# Plasmon Resonance in Multilayer Graphene Nanoribbons

*Naresh Kumar Emani[1,3], Di Wang[1,3], Ting-Fung Chung[2,3], Ludmila J. Prokopeva[1,4], Alexander V. Kildishev[1,3], Vladimir M. Shalaev[1,3], Yong P. Chen[2,1,3], and Alexandra Boltasseva[1,3,5]*

*Corresponding Author: E-mail: aeb@purdue.edu

[1]School of Electrical and Computer Engineering, Purdue University, West Lafayette, IN 47907, USA
[2]Department of Physics and Astronomy, Purdue University, West Lafayette, IN 47907, USA
[3]Birck Nanotechnology Center, Purdue University, West Lafayette, IN 47907, USA
[4]Novosibirsk State University, Novosibirsk, Russia
[5]DTU Fotonik, Department of Photonics Engineering, Technical University of Denmark, Lyngby, DK-2800, Denmark

ABSTRACT Plasmon resonance in nanopatterned single layer graphene nanoribbon (SL-GNR), double layer graphene nanoribbon (DL-GNR) and triple layer graphene nanoribbon (TL-GNR) structures is studied both experimentally and by numerical simulations. We use 'realistic' graphene samples in our experiments to identify the key bottle necks in both experiments and theoretical models. The existence of electrical tunable plasmons in such stacked multilayer GNRs was first experimentally verified by infrared microscopy. We find that the strength of the plasmonic resonance increases in DL-GNR when compared to SL-GNRs. However, we do not find a further such increase in TL-GNRs compared to DL-GNRs. We carried out systematic full wave simulations using finite element technique to validate and fit experimental results, and extract the carrier scattering rate as a fitting parameter. The numerical simulations show remarkable agreement with experiments for unpatterned SLG sheet, and a qualitative agreement for patterned graphene sheet. We believe that further improvements such as introducing a bandgap into the numerical model could lead to a better quantitative agreement of numerical simulations with experiments. We also note that such advanced modeling would first require better quality graphene samples and accurate measurements.

# 1. Introduction

Graphene has emerged as a versatile and dynamic platform for hybrid nanophotonics and optoelectronics due to its excellent electrical and optical properties[1-4]. This material has recently been integrated with metamaterials[5], plasmonic nanoanteannas[6-11], waveguides[12] and photonic crystals[13, 14] to realize electrically tunable hybrid devices. Nanostructured graphene has been shown to support highly confined surface plasmons with plasmon wavelength being 40-100 times smaller than free space wavelength at mid-infrared wavelengths[15-18]. These plasmon modes in graphene can be electrically controlled and have tremendous potential for confining and manipulating radiation for mid-infrared applications[15-17, 19, 20]. At present there are two main challenges in the area of graphene plasmonics: to drive the plasmonic resonance to near-infrared wavelengths, and to increase the relatively small strength of the plasmon resonance which is due to finite optical conductivity of single layer graphene (SLG)[21]. Optical studies of AB-stacked bilayer graphene using a synchrotron light source reveal that the optical conductivity of multilayer graphene is higher than SLG[22]. While the optical conductivity of SLG is consistent with the prediction of the Random Phase Approximation (RPA) theory[4, 23], the spectrum of AB-stacked bilayer graphene shows a sharp resonance at 0.37 eV due to interlayer coupling[22]. Theoretical studies also predict such an enhanced optical conductivity in bilayer graphene due to strong interlayer coupling[24]. If the number of graphene layers is further increased the optical conductivity spectrum becomes progressively more complex, but the general trend of increasing optical conductivity is maintained[22]. On the other hand carrier mobility, which determines loss of the plasmonic resonance, decreases when the number of layers is increased due to modification of the electronic bandstructure. Until now the studies in graphene plasmonics have focused on SLG which can be synthesized into large area samples reasonably easily. Due to enhanced optical conductivity, multilayer graphene could support stronger plasmonic resonance when compared to SLG. Further in multilayer graphene a perpendicular electric field could be applied to achieve stronger control on plasmonic resonance[25]. In this paper we present our experimental and numerical studies on plasmon resonance in 'realistic' randomly-stacked multilayer chemical vapor deposition (CVD) grown graphene nanoribbons (GNRs).

## 2. Experiment

We transfer and stack CVD grown SLG sheets to form multilayer graphene samples due to difficulties in obtaining large area samples with controlled number of layers by mechanical exfoliation. SLG was first grown on 25 *µm*-thick Cu foils using an atmospheric pressure CVD process[26-28]. It was then sequentially transferred assisted by poly(methyl methacrylate) (PMMA)[26, 29] on three separate silicon substrates (1-10 Ohm-cm) with 300 nm thermal oxide (Si/SiO$_2$) to form single layer, double layer, and triple layer devices. Subsequently a 500 µm × 500 µm active area was defined by photolithography and oxygen plasma etching. The source-drain contacts were defined by photolithography and subsequent Ti and Au metallization (5 nm and 55 nm respectively) on each sample.

It is well known that the layer stacking order in multilayer graphene plays a crucial role in determining its optical properties[30]. In our samples the domain orientation is not uniform across the graphene layer and there is also no definite stacking order between adjacent layers. Hence the optical response will be averaged over many domains with random orientations in a large area. We performed Raman spectroscopy (532 nm, circularly polarized laser with ~1 µm spot size and 1 mW incident power on the sample) to probe local layer orientations in our samples, since it has been shown to be a sensitive probe of the unique electronic and phonon band structures in graphene layers[31]. From the Raman spectra (shown in Figure 1 (a-c)) we clearly observe the $I_{2D}/I_G$ ratio is dependent on the measurement location in two layer and three layer graphene samples in contrast to the single layer sample. This is due to changes in local lattice stacking order, which is consistent with previous studies in misoriented graphene[30, 32]. Hence, we should note that there is significant inter-layer stacking misalignment in addition to the well-known intra-layer domain misalignment in CVD graphene samples. Electrical testing of the devices was also carried out to verify the gate modulation of the source drain sheet resistance in multilayer graphene sheets. We found that SLG exhibits the highest dynamic range of variation of electrical resistance as shown in Figure 1(d), followed by two layer graphene and three layer graphene respectively. The gate induced carrier density modulation will allow multilayer GNRs to support tunable plasmonic resonance. The strength of such resonance would be strongly dependent on the optical conductivity and carrier mobility.

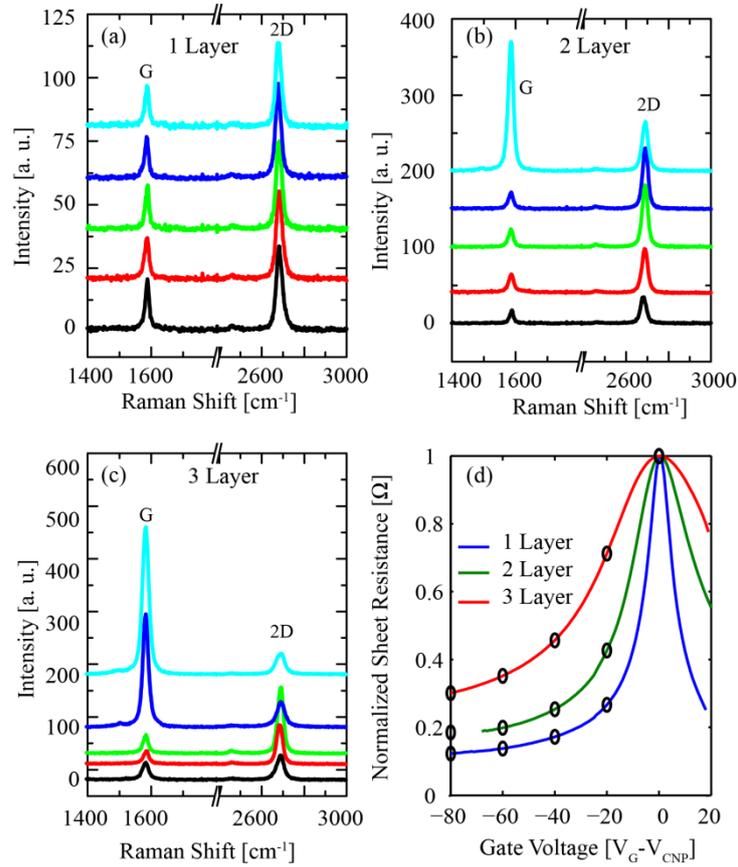

**Figure 1.** (a-c) Raman spectra collected from five random locations on single layer, 2 layer and 3 layer graphene respectively. All the measurements were performed using a 532 nm, circularly polarized laser source with a 100X objective (spot size ~1 μm) and 1 mW incident power. Individual spectra are offset for clarity. (d) The gate modulation of source drain resistance (normalized using sheet resistance at the charge neutral point (CNP) voltage) in different samples. SLG exhibits highest dynamic range of variation of electrical resistance

followed by 2 layer and 3 layer graphene respectively. The open circles represent the gate voltages at which IR reflection data shown later is collected.

The active area was patterned into GNRs (50-nm width and 150-nm period) using electron beam lithography on a positive ebeam resist (ZEP 520A, Zeon Chemicals, Inc). Figure 2 shows a simplified schematic illustration of our experimental setup as well as a scanning electron micrograph showing the patterned graphene ribbons. The number of broken C-C bonds increases significantly in nanopatterned graphene in comparison to unpatterned large area graphene, which leads to an additional peak ($\sim 1350$ cm$^{-1}$) in the Raman spectra (see SOM Figure S4 for an example for SLG). However, we should note that even after patterning the $I_{2D}/I_G$ ratio of SLG is greater than 2, indicating that the physical properties of graphene are intact. To investigate the plasmonic resonance in GNRs we measure the IR reflectance which we normalize to the reflectance at charge neutral point (CNP). The optical measurements were performed using a Fourier Transform Infrared (FTIR) spectrometer (Nicolet Magna-IR 850) with a microscope accessory (Nicplan IR Scope, 15X, NA 0.58 Reflectochromat objective). The incoming beam was polarized with electric field perpendicular to ribbons using a wire grid polarizer to excite transverse magnetic modes in GNRs.

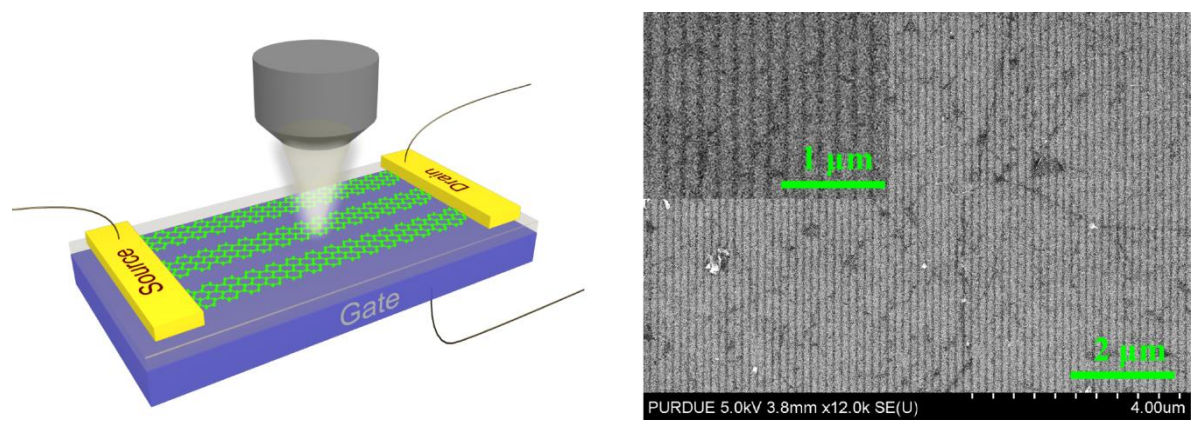

**Figure 2.** (a) Simplified schematic of the experimental setup used for studying plasmon resonance in GNRs. The lattice orientation of GNRs in the figure is for illustration only and dimensions are not to scale; (b) Scanning electron micrograph of the fabricated GNRs on SLG sample with the inset showing a zoomed-in view of GNRs.

When graphene is patterned into nanoribbons it can support surface plasmon standing waves when the condition $\text{Re}(\beta)W = m\pi + \phi$ is satisfied, where $\beta$ is the surface plasmon propagation constant, $W$ is the width of the GNR, and $\phi$ is an arbitrary phase shift introduced by the reflection at the GNR edge and m is an integer[15, 18, 33]. Plasmon resonances in 50-nm-wide GNRs occur in the wavelength range of 7 $\mu m$ – 10 $\mu m$ when graphene is doped to $1\times10^{12}$ – $7\times10^{12}$ $cm^{-2}$ carrier densities. The experimental measurements of normalized reflectance on SL-GNRs, DL-GNRs and TL-GNRs as a function of Fermi energy $E_F$ (which is related to carrier density, see supplementary online material (SOM) Section VI) are shown in Figure 3. As the carrier density in GNRs is increased the plasmon resonance becomes stronger, and the resonance moves to lower wavelengths. There are two main peaks observed in the measured data – one above and another below the optical phonon wavelength of $SiO_2$. These peaks result from hybridization of graphene plasmon with the optical phonon in the $SiO_2$ layer[15, 18, 34]. The resonance strength increases from SL-GNR to DL-GNR, but is slightly weaker for TL-GNRs. This could be due to higher losses resulting from the increase in the number of defects arising out of stacking multiple layers.

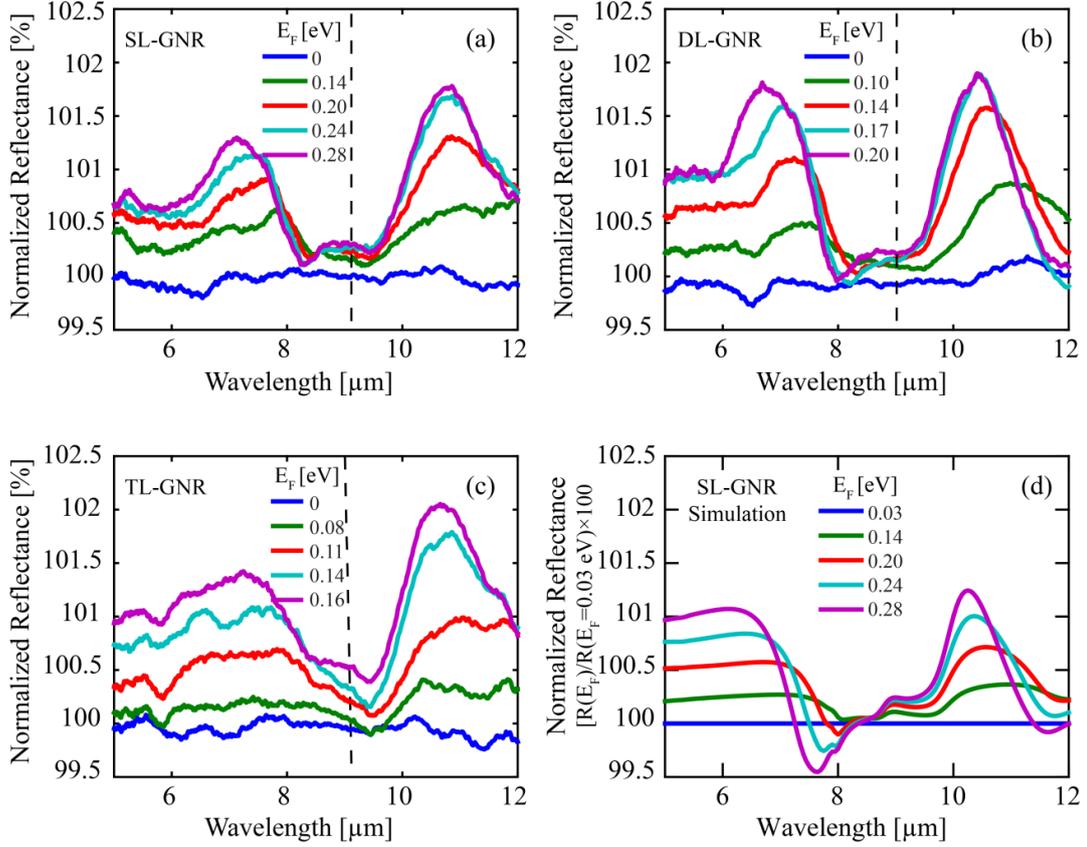

**Figure 3.** Modulation of IR reflectivity of GNRs fabricated on Si/SiO$_2$ substrate as a function of Fermi energy ($E_F$) of graphene; the vertical dashed line indicates the peak of SiO$_2$ optical phonon. Panels (a) - (c) show measured data on SL-GNRs, DL-GNRs and TL-GNRs respectively. The reflection measurements were normalized to the reflection at the charge neutral point in our experiments. The width and period of GNRs were fixed at 50 nm and 150 nm respectively. (d) 2D full wave FEFD simulations of SL-GNRs with COMSOL Multiphysics using a surface current model for graphene; simulations performed at 0° to 35° angles of incidence ($\varphi$) with 5° spacing were averaged to obtain the curves shown here (see SOM on substrate characterization for further details). The Fermi energy for each sample was calculated using a uniform charge approximation which does not take into account the screening and interlayer coupling effects (see SOM section VI for further details).

In Figure 4 we plot the peak intensities of the resonance peaks shown in Figure 3 as a function of Fermi energy. We find that the peak resonance intensities in DL-GNRs are significantly stronger than SL-GNRs at a fixed $E_F$. When the $E_F$ is held constant the total carrier concentration of the stack is simply the carrier concentration in SL-GNRs times the number of graphene layers. We should note that a similar strong increase in peak intensity of TL-GNRs is not seen when compared to DL-GNRs. We believe that this could be due to the fact that the PMMA assisted transfer of CVD graphene invariably creates some holes, folds and unavoidable residue. In fact, as we increase the number of layers the non-uniformities become quite apparent during SEM imaging (see Figure 2b for a representative image of SL-GNRs) and under an optical microscope. Therefore, we believe that this increase in number of defects per unit area leads to progressively higher losses, and weaker response which manifests as broadening of plasmon peak in Figure 3(a-c). While the quality of our samples is comparable to the current state of the art in CVD graphene, we can expect that further improvements in graphene growth/transfer processes will help in further enhancing the plasmon resonance strength.

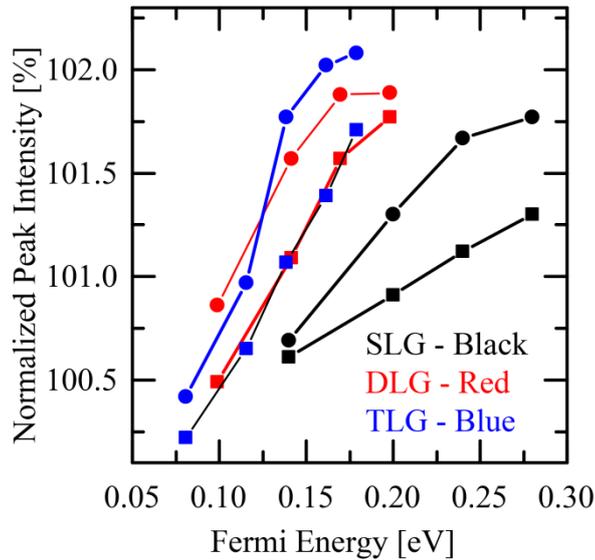

**Figure 4.** Peak intensity of the resonances peaks shown in Figure 3 as a function of $E_F$. The graphene plasmon hybridizes with the $SiO_2$ optical phonon to give two peaks shown in square and circle markers respectively. Square markers indicate resonance peaks at shorter wavelengths, while circles indicate resonance peaks at longer wavelengths.

**3. Numerical Simulations**

To gain further insight into the experiments we performed full wave finite element frequency domain (FEFD) simulations using a commercial software package (COMSOL Multiphysics, Wave Optics Module). We first accurately determined the dielectric function of $SiO_2$, which has a strong optical phonon overlapping with the graphene plasmon[15, 34], using IR spectroscopic ellipsometry. The retrieved optical constants for Si and SiO$_2$ layers are used in subsequent simulations and are shown in Figure S1. The optical properties of graphene were calculated using the local limit of the Random Phase Approximation (RPA) and were modelled as a surface current in FEFD simulations.

The surface current model in COMSOL was first validated for an unpatterned single graphene sheet on SiO$_2$/Si substrate by modifying the classical Drude equation for the complex reflection coefficient[35] now rewritten as $r = \left( r_{01}^{\pm} + r_{01}^{\mp} r_{12} e^{\iota 2 k_1 \delta} \right) / \left( 1 + r_{01}^{=} r_{12} e^{\iota 2 k_1 \delta} \right)$. The classical Fresnel coefficient for the *p*-polarized light, $r_{12} = \left( \varepsilon_2 k_1 - \varepsilon_1 k_2 \right) / \left( \varepsilon_2 k_1 + \varepsilon_1 k_2 \right)$, was still applicable at the second interface with no graphene sheet, with $k_i = \frac{\omega}{c}\sqrt{\varepsilon_i - \varepsilon_0 \sin^2 \varphi}$, $i \in \overline{0,2}$ for a given frequency of light $\omega$ and angle of incidence $\varphi$. While using $\xi = \sigma k_0 k_1 / \omega \varepsilon_0$ three different permutations of a modified Fresnel coefficient at the first interface, $r_{01}^{\pm} = \left( \varepsilon_1 k_0 + \xi - \varepsilon_0 k_1 \right) / \left( \varepsilon_1 k_0 + \xi + \varepsilon_0 k_1 \right)$, $r_{01}^{\mp} = \left( \varepsilon_1 k_0 - \xi + \varepsilon_0 k_1 \right) / \left( \varepsilon_1 k_0 + \xi + \varepsilon_0 k_1 \right)$, and $r_{01}^{=} = \left( \varepsilon_1 k_0 - \xi - \varepsilon_0 k_1 \right) / \left( \varepsilon_1 k_0 + \xi + \varepsilon_0 k_1 \right)$, were required to account for the effect of the graphene layer. Here, $\varepsilon_0, \varepsilon_1, \varepsilon_2$ are the dielectric constants of air, SiO$_2$ and Si substrate, and $\sigma$, $\delta$, and $c$ are the conductivity of the graphene layer, the thickness of silicon dioxide and the free-space speed of light respectively. Further details of our implementation and can be found in SOM.

We found that simulations at only normal incidence do not fully account for all the experimental features (see SOM Figure S2). Therefore we developed a weighted averaging procedure where contribution of each simulation performed with 0° to 35° angles of incidence was weighted with a Gaussian factor. The upper limit of 35° was chosen to account for the finite acceptance angle of the objective used in our experiment. The final results thus obtained capture the experimental data remarkably well as compared to just normal reflectance as shown in Figure S2. From this analysis we retrieved a carrier scattering time of ~10 *fs* for the unpatterned graphene sample which is 5 times lower than the value estimated using DC Drude model[3] (see SOM Section II and VI for additional details). We also recently became aware of another work which reports an experimentally extracted scattering time of 18 *fs,* which is in the same range as our results[36]. In numerical simulations SL-GNRs were modelled as patterned surface current. The results obtained with the averaging procedure described above are shown in Figure 3(d), where we see a qualitative agreement with the experimental results. A key difference is the considerably narrower plasmon peaks below the $SiO_2$ optical phonon wavelength in experiments when compared to simulations. When graphene is patterned into nanoribbons the carriers are confined to a 1D strip leading to opening of an energy bandgap. At the same time there is also significant edge disorder leading to charge localization and a smaller effective width of the GNR[37]. The bandgap ($E_{gap}$) is found to be empirically related to GNR width (W) and disorder parameter ($W^*$) as $E_{gap} = \alpha/(W-W^*)$ based on electrical transport studies on epitaxial graphene[37]. According to these studies a rather large bandgap of 0.2 eV can be obtained for GNR widths of ~15 nm. It seems plausible that such a bandgap could reduce the optical loss at IR wavelengths, and consequently lead to narrower plasmon resonance peaks. While we can expect significant differences between electrical and optical responses, it seems plausible that such a bandgap could reduce the optical loss at IR wavelengths. Based on our numerical studies we conclude that the experimental features cannot be attributed to variations in the width of the ribbons or carrier scattering time alone. Therefore, we believe that the optical conductivity for graphene ribbons should be re-derived taking into account the energy bandgap which is beyond the scope of this work.

## 4. Conclusions and outlook

A major current challenge in the area of graphene plasmonics is to improve the strength of the plasmonic resonance. CVD grown graphene, which yields large sample area, has been predominantly used in graphene plasmon studies due to ease of optical characterization.

However, growth kinetics and transfer method of CVD graphene lead to disorder and hence poorer physical properties compared to epitaxial graphene films on silicon carbide. We investigated the behavior of plasmon resonance in GNRs in single layer and multi-layer 'realistic' CVD graphene. Our experimental results indicate that plasmons are indeed supported by multilayer graphene nanostructures. When the carrier concentration of the graphene sheet is fixed, we find that DL-GNRs show stronger plasmon peak when compared to SL-GNRs. However, the strength of plasmon peak did not further increase from DL-GNRs to TL-GNRs most likely due to inhomogeneities in local stacking order as well as random orientation of domains within CVD graphene. Systematic numerical simulations were performed in order to obtain a very good fit with experimental results for unpatterned graphene. Thus, we retrieved a carrier scattering time of ~10 $fs$ from our graphene sample and developed an accurate numerical model which takes into account contributions from 0° to 35° incidence angles. The developed simulation model was applied for GNRs, and the results agree qualitatively with the experiment, but show broader plasmonic resonances. We believe that this could be due to opening of the bandgap close to the Dirac point due to nano-patterning. While incorporating a bandgap into the numerical model could theoretically lead to a better fit, we believe that such advanced modelling would first require better quality graphene samples and accurate measurements.

**Supporting Information**

Additional discussion on optical characterization of the substrate, modelling of unpatterned graphene, convergence problems with finite thickness model of graphene, Raman spectroscopy of GNRs and calculation of the Fermi energy, Drude scattering rate and mobility is provided as supplementary online material (SOM).

**Funding Sources**

The authors want to acknowledge financial support from ARO MURI Grant 56154-PH-MUR (W911NF-09-1-0539) and NSF Materials Research Science and Engineering Center (MRSEC) program DMR1120923. The graphene synthesis and Raman characterization were supported by NSF DMR 0847638.

**Acknowledgements**

The authors thank Dr. Tom Tiwald (J. A. Woollam Co.) for performing the IR spectroscopic ellipsometry of the substrates used in this work, and a very useful discussion on substrate phonon characterization. We also thank Clayton DeVault for assistance with preparation of the manuscript.

**Abbreviations**

CVD – Chemical Vapor Deposition, SLG – Single Layer Graphene, GNRs – Graphene Nanoribbons, SL-GNRs – Single Layer GNRs, DL-GNRs – Double Layer GNRs, TL-GNRs – Triple Layer GNRs, CNP – Charge Neutral Point, FTIR – Fourier Transform Infrared Spectroscopy

**Keywords**: Graphene plasmonics, multilayer CVD graphene, plasmon-substrate phonon interaction

Supporting Information

# Plasmon Resonance in Multilayer Graphene Nanoribbons

*Naresh Kumar Emani[1,3], Di Wang[1,3], Ting-Fung Chung[2,3], Ludmila J. Prokopeva[1,4], Alexander V. Kildishev[1,3], Vladimir M. Shalaev[1,3], Yong P. Chen[2,1,3], and Alexandra Boltasseva[1,3,5*]*

### I. Optical Characterization of the substrate

We chose a silicon substrate with low doping (1-10 Ohm-cm) to avoid additional artifacts due to free carrier absorption in highly doped silicon substrates. A thermally grown 300 nm oxide serves as a gate dielectric. We used IR spectroscopic ellipsometry[1] (J. A. Woollam Co) to accurately characterize the optical properties of $SiO_2$ layer on top of Si, especially around the optical phonon in $SiO_2$ which is strong between the 8 µm – 10 µm wavelength range. The measured data was fitted with 9 Gaussian oscillators to obtain the dielectric function of $SiO_2$.

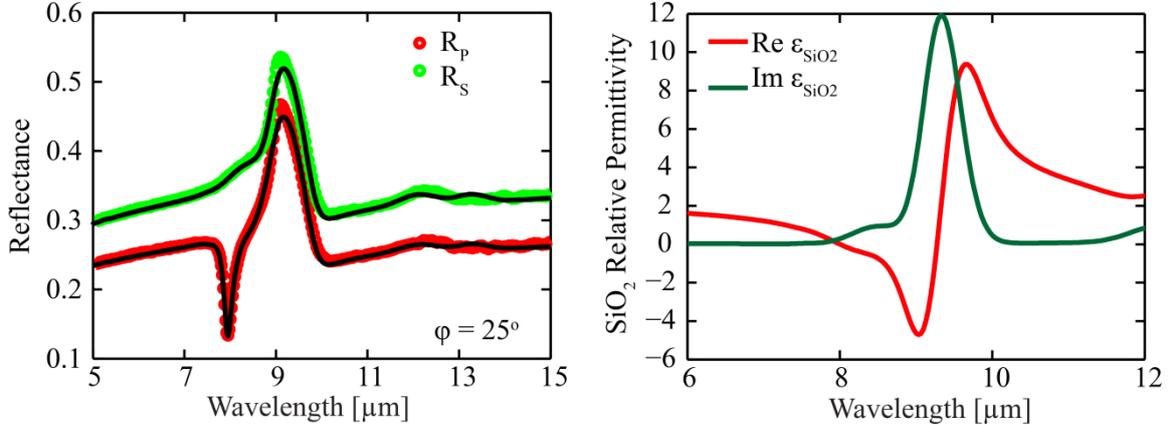

**Figure S1**. (a) Spectroscopic ellipsometry measurements of lightly doped Si with 300 nm thermal oxide at angle of incidence $\varphi = 25°$. The black solid lines show the numerical fit of measured data using 9 Gaussian oscillators. The refractive index of Silicon was extracted to be a constant value of 3.42 in this wavelength range; (b) The extracted permittivity around the $SiO_2$ optical phonon wavelength.

---

[1] We thank Dr. Tom Tiwald at J A Woollam Co. for performing these measurements, and helping with retrieval of accurate optical constants of $SiO_2$.

## II. Characterization of the single layer graphene (SLG) sheet on the substrate

The optical measurements of bare (unpatterned) graphene sheet put on the same substrate (air/300-nm SLG-covered $SiO_2$ film/semi-infinite Si substrate) are shown in Figure S2(a). The reflectance data is pinned in the 9 - 10 μm wavelength range due to the strong optical phonon. There is an additional spectral feature at 8 μm that appears in the measurements.

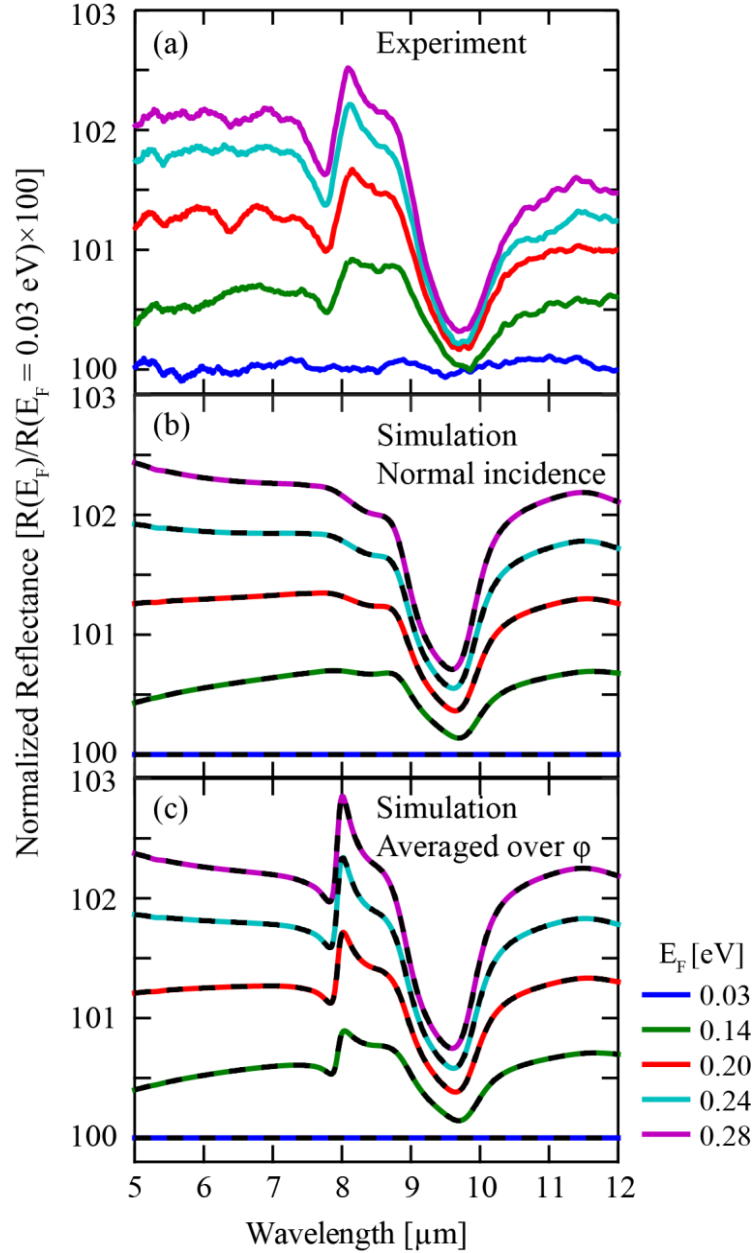

**Figure S2**. (a) Modulation of IR reflectivity of CVD SLG on $Si/SiO_2$ substrate as a function of wavelength for different values of Fermi energy $E_F$. Measurements were performed using a Fourier Transform Infrared (FTIR) Spectrometer using unpolarized light and a microscope

accessory (Objective: 15X, N.A. 0.58 Reflectochromat); measurements are normalized to reflection at the charge neutral point; (b)-(c) Analytical simulation results obtained at normal incidence and by averaging multiple simulations with Eq. 1 over a range of angle of incidence ($\varphi$) respectively. We extracted a carrier scattering time of 10 *fs* from our numerical analysis. The dashed black curves show the matching results of the FEFD calculations that were used to validate the surface current approach with SLG models.

In simulations, we introduced graphene to the model with a sheet surface current ($J = \sigma E$) at the air/SiO$_2$ interface where the $\sigma$ is the surface conductivity of graphene derived within the local limit of Random Phase Approximation[1,2] (RPA) and $E$ is the incident electric field. *The carrier scattering rate* was used *as a fitting parameter* in numerical simulations and it was extracted to be approximately $1\times10^{-14}$ s. The RPA model in conjunction with accurate substrate optical constants put in the *modified Drude equation* (see section below) captures the main experimental trends very well as shown in Fig. S2(b) for normal incidence. However, there are sharp features at a wavelength of 8 μm, which can potentially interfere with graphene plasmons in the same wavelength range. Our analysis reveals that this is due to zero crossing of the real part of dielectric function of SiO$_2$ at exactly 8 μm wavelength. At the so-called epsilon near zero (ENZ) wavelength the absorption in SiO$_2$ is enhanced for *p*-polarized light due to concentration of electric field in a region with non-zero imaginary part of dielectric function[3]. This ENZ enhanced absorption is strongly dependent on angle of incidence and vanishes at normal incidence. Due to imperfections and finite numerical aperture (NA) of the Reflectochromat we have contributions from off-normal angles of incidence. The exact contribution of each angle is difficult to determine, therefore we chose to average simulations over a range of angles of incidence (0° - $\sin^{-1} NA \sim 35°$ with a 5° step size). The simulation result from each angle of incidence ($\varphi$) was weighted with the Gaussian distribution function $\mathrm{e}^{-(\varphi/35°)^2}$ to capture the experimental data better. We dropped multiplicative factor of distribution because the averaged values for different Fermi energies were then normalized by that at CNP. The final result of our averaging procedure shows remarkable agreement with experimental data as shown in Fig. S2(c), as opposed to just normal simulation in Fig. S2(b).

We also performed Finite Element Frequency Domain (FEFD) test simulations in COMSOL Multiphysics to validate the surface current approach. Results demonstrate absolute and exact match with analytics formulas (see dashed lines in Fig. S2 (bc)) without any spectrum shifts typical for thickness-based simulations (see section below). Our verified surface current model and averaging procedure were used to obtain the simulation curves for SL-GNR presented in Fig. 4, where analytical formulas were not available.

### III. Modified Drude equation for the reflection coefficient of the SLG-covered dielectric film

The classical Drude equation[4] for the complex reflection coefficient $r$ of a film (of thickness $\delta$ and permittivity $\varepsilon_1$) deposited on a semi-infinite substrate for $p$-polarized light can be modified in order to account for an SLG layer on top of a film

$$r = \frac{r_{01}^{\pm} + r_{01}^{\mp} r_{12} e^{i 2 k_1 \delta}}{1 + r_{01}^{=} r_{12} e^{i 2 k_1 \delta}}, \quad R = |r|^2, \tag{1}$$

where three distinct permutations of a modified Fresnel coefficients with a normalized quantity $\xi = \sigma k_0 k_1 / \omega \varepsilon_0$ are required to describe the effect of the SLG at the superstrate-to-SLG-covered film interface, i.e. $r_{01}^{\pm} = \varepsilon_1 k_0 + \xi - \varepsilon_0 k_1 \,/\, \varepsilon_1 k_0 + \xi + \varepsilon_0 k_1$, $r_{01}^{\mp} = \varepsilon_1 k_0 - \xi + \varepsilon_0 k_1 \,/\, \varepsilon_1 k_0 + \xi + \varepsilon_0 k_1$, and $r_{01}^{=} = \varepsilon_1 k_0 - \xi - \varepsilon_0 k_1 \,/\, \varepsilon_1 k_0 + \xi + \varepsilon_0 k_1$, while, $r_{12} = \varepsilon_2 k_1 - \varepsilon_1 k_2 \,/\, \varepsilon_2 k_1 + \varepsilon_1 k_2$ is the classical Fresnel coefficient at the film-to-substrate interface with no graphene sheet. Here, $k_i = \frac{\omega}{c}\sqrt{\varepsilon_i - \varepsilon_0 \sin^2 \varphi}$, $i \in \overline{0,2}$ for a given frequency of light $\omega$ and angle of incidence $\varphi$ and $\varepsilon_0, \varepsilon_2$, $\sigma$, and $c$ are the dielectric constants of superstrate, substrate, the conductivity of the graphene layer, and the free-space speed of light respectively.

This reflection coefficient formula can be rewritten in *more compact and cascading friendly form* using material matrix notation for each $i$ to $j$ interface with (possibly zero) surface conductivity $\sigma_{ij}$

$$\mathbf{m}^{ij} = \frac{\varepsilon_i}{k_i}\begin{bmatrix} 1 & -1 \\ -1 & 1 \end{bmatrix} + \frac{\sigma_{ij}}{\omega \varepsilon_0}\begin{bmatrix} 1 & -1 \\ 1 & -1 \end{bmatrix} + \frac{\varepsilon_j}{k_j}\begin{bmatrix} 1 & 1 \\ 1 & 1 \end{bmatrix}, \; ij \in \{01, 12\}. \quad (2)$$

Reflection coefficient with this notation reads

$$r = \frac{m_{21}^{01} + m_{22}^{01}\rho}{m_{11}^{01} + m_{12}^{01}\rho}, \; \rho = \frac{m_{21}^{12}}{m_{11}^{12}} e^{\iota 2 k_1 \delta}, \quad (3)$$

In this form it can be further straightforwardly cascaded to obtain reflection coefficient for any multilayer structure with possibly conductive interfaces between layers.

## IV. Convergence of the finite-thickness model for graphene nanoribbons

Atomic-scale thickness of graphene sheets can additionally complicate accuracy and efficiency of the numerical simulations with multilayer graphene and Graphene Nanoribbons (GNR). So far, an intuitive and most popular way of introducing multivariate RPA surface conductivity $\sigma(\omega)$ to conventional 3D computational electromagnetics (CEM) solvers has been to introduce an artificial finite thickness ($\delta_g$) and a corresponding volume permittivity ($\varepsilon = 1 + \iota \sigma \left( \omega \delta_g \right)^{-1}$) for graphene elements. However, the thickness should be small enough to guarantee the convergence to the surface conductivity model, typically, on the order of 1 nm for an unpatterned SLG. For nanostructured graphene, requirement on $\delta_g$ can be two orders of magnitude smaller[5, 6]. Volume implementation of the graphene elements can lead to inexact and enormously expensive computations because of an extremely fine computational mesh (sub-gridding helps only partially), fictitious volume modes, spectral shifts, poor convergence of modal methods etc. The computational complexity grows at least quadratically (2D) or cubically (3D) with the linear mesh size, so that even single runs can become computationally expensive, making fitting (for material parameter retrieval) or optimization tasks almost impossible. However, all the above complications are introduced artificially and can be avoided just by using native, surface-based numerical models. In this paper, we use a mesh-based commercial solver, where we introduce a GNR into the model as a surface current. Thus, we avoid errors from finite thickness approximation and we avoid extra-fine meshing. Our modification of other popular mesh-based and meshless CEM solvers will be published elsewhere.

In figure S3, we show the results of thickness convergence analysis for SL-GNR and. The normalized reflectance for 35° angle of incidence and 0.28 $eV$ Fermi energy is shown for different graphene thicknesses 0.1, 0.5, 1, and 2 nm. The dashed line present reflectance obtained with surface FEFD model. We observe that the different thickness curves are somewhat similar, yet complete stabilization is achieved only for a thickness of 0.1 nm as can be seen in Fig. S3(b). For larger thicknesses reflectance is red shifted, peaks are enhanced and broadened. In cascaded GNR (not shown here), as e.g. in a graphene based pulse shaping device[5], the reflectance is more sensitive to insufficiently small thicknesses and this effect is much more pronounced with one order smaller required thickness. Relatively small artificial thickness discrepancies may not always be a problem for a brief visual comparison to experiments, however they introduces large function difference in the areas with large gradients causing problems to fitting procedure and parameter retrieval.

In conclusion, the errors in simulations from the finite thickness of graphene are critical especially for fitting procedures. For best computational practices, surface models shall be used.

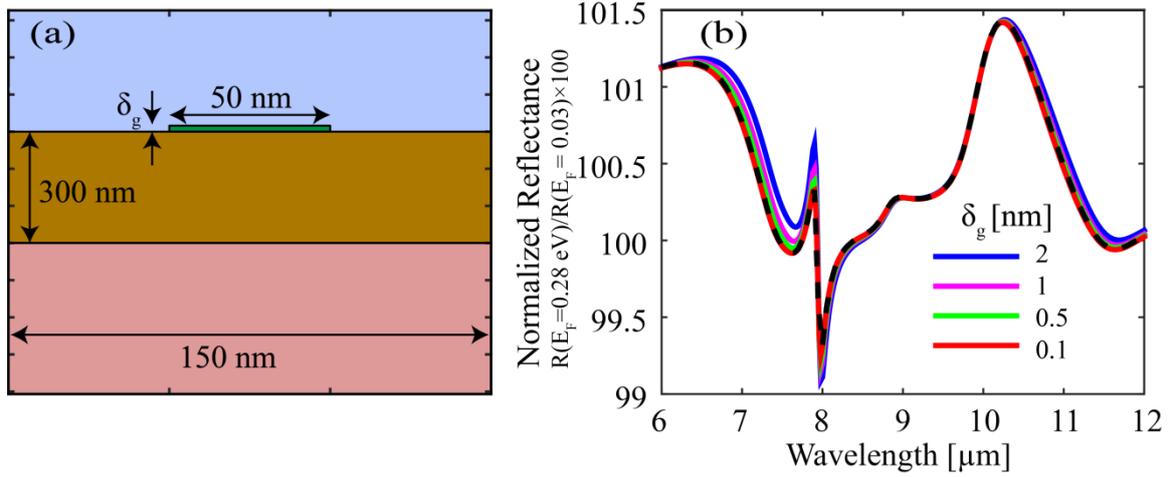

**Figure S3** (a) Schematics of SL-GNR sample used in our experimental and numerical studies; (b) Convergence of normalized reflectance in SL-GNR structures simulated using volume implementation of graphene with fictitious thicknesses ($\delta_g$); dashed line shows results of surface model.

## V. Raman Spectroscopy of Graphene Nanoribbons

The presence of disorder in graphene will lead to a peak in Raman spectra at $\simeq 1350$ cm$^{-1}$ which is referred to as D peak[7, 8]. Raman spectroscopy was performed to verify the extent of defects in bare graphene and GNRs using a Horiba Jobin Yvon Xplora confocal Raman microscope. The analysis of different peaks (parameters listed in table S2) reveals that patterned graphene retains the physical properties of graphene sheet with aa significantly enhanced D peak at 1343 cm$^{-1}$ due to structural defects such as graphene edges[7]. Further, the $I_{2D}/I_G$ ratio which is indicative of quality of graphene is greater than 2.

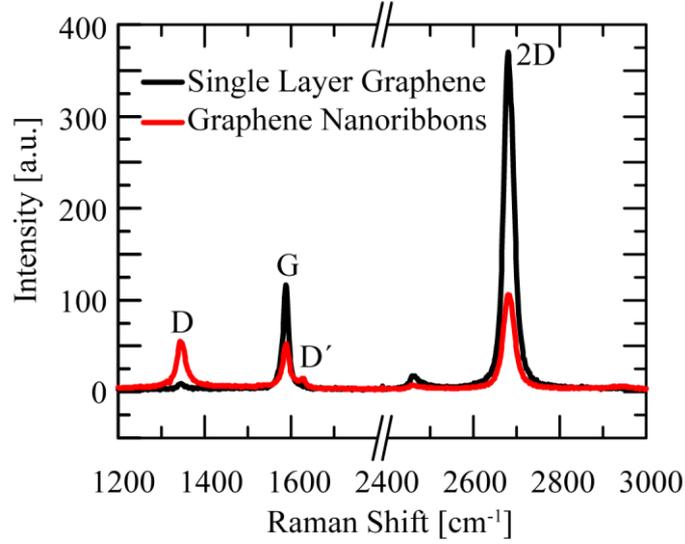

**Figure S4**. Raman spectra of unpatterned SLG and GNRs. All measurements were performed under identical conditions with 532 nm, 1 mW, circularly polarized laser source and 100X objective. Resolution of spectrometer is 1.3 cm$^{-1}$. Patterning causes a significant enhancement in the D peak intensity while simultaneously decreasing the $I_{2D}/I_G$ ratio from 3.18 to 2.08.

|           | Frequency [cm$^{-1}$] | FWHM [cm$^{-1}$] | Intensity [a. u.] |
|-----------|------------------------|-------------------|--------------------|
| G (SLG)   | 1588                   | 15                | 116                |
| 2D (SLG)  | 2680                   | 28                | 370                |
| G (GNRs)  | 1589                   | 19                | 50                 |
| 2D (GNRs) | 2680                   | 15                | 104                |
| D (GNRs)  | 1343                   | 25                | 53                 |

**Table 1:** Analysis of the Raman spectra for unpatterned graphene and GNRs. Peak frequencies, full width at half maximum (FWHM) and intensities are extracted by fitting these peaks with Lorentzian function.

## VI. Estimation of carrier density, Fermi energy and Drude scattering time

The carrier density in our samples was estimated using a simple parallel plate capacitor model given by $n_{graphene} = C_{gate}(V_G - V_{CNP})/q$, where $C_{gate} = \frac{\varepsilon_{oxide}}{t_{oxide}} = 11.5 \frac{\text{nF}}{\text{cm}^2}$ is the gate capacitance, V$_G$ is the applied gate voltage and V$_{CNP}$ is the charge neutral point voltage and q is the charge of the electron. The Fermi energy ($E_F$) of SLG was calculated assuming a linear dispersion model which results in $E_F = \hbar v_F \sqrt{\pi n_g}$. For multilayer samples we assumed that the charge is uniformly distributed among K layers giving $E_F = \hbar v_F \sqrt{\pi n_g / K}$, where K is the number of graphene layers. An alternative method to estimate the charge in each layer would be to consider the total charge supported by gate and then calculate the screened charge in each layer using the equation: $Q_{Gate} = Q_1 + Q_2 + Q_3 = Q_1[1 + \exp(-d/\lambda) + \exp(-2d/\lambda)]$, where $Q_{Gate}$, $Q_1$, $Q_2$ and $Q_3$ are charge densities supported by gate, 1st layer, 2nd layer and 3rd layer respectively, d is the interlayer distance and $\lambda$ is the Thomas-Fermi screening parameter [9]. In our analysis we use a uniform charge distribution approximation, which we believe is valid up to a few layers, for three main reasons:

1. The interlayer coupling and Thomas Fermi screening parameters are well known for epitaxial multilayer graphene but not stacked CVD graphene used in our experiments. In case of epitaxial bilayer graphene the charge distribution ratio would be 1:0.65 (assuming $\lambda$ =0.7 nm and d=0.3 nm[9]). Further, the interlayer spacing is not uniform in our stacked sample as revealed by AFM measurements. Therefore, in the absence of reliable parameters we feel justified in making a uniform distribution approximation.

2. In optical experiments, the multilayer graphene sheet will behave as an effectively uniform charge sheet due to extreme subwavelength length scale.

3. As a confirmation of our hypothesis, our analysis shows that even simple local-carrier models give a good agreement after accounting for off-normal contributions of the polar substrate.

Using the data shown in Fig 1d we extract the mobility and DC scattering time for SLG device on which optical characterization (SOM Section II) was performed. The extracted mobility is comparable to the literature on CVD graphene devices. However, surprisingly the Drude scattering time is only around 40-50 *fs,* and this is about 5 times larger than the time retrieved from our optical measurements. Therefore the material properties of graphene should be chosen after careful consideration so as to reflect the real experimental devices.

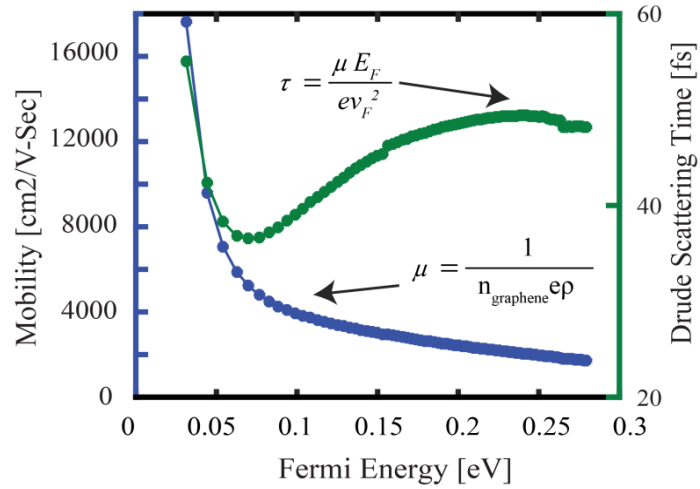

**Figure S5**. Mobility and Drude scattering time estimated ($v_F$ =10$^8$ *cm/s*) from the electrical characterization shown in Fig 1(d).